\documentclass{jetpl}
\twocolumn

\lat

\title{The possibility of Z(4430) resonance structure description in
 $\pi\psi'$ reaction}

\rtitle{}

\sodtitle{On the possibility to find Z(4430) resonance structure
in $\pi\psi'$ reaction}

\author{I.\,V.\,Danilkin$^{+*}$\/\thanks{e-mail: danilkin@itep.ru},
P.\,Yu.\,Kulikov$^+$} 

\rauthor{I.\,V.\,Danilkin, P.\,Yu.\,Kulikov }

\sodauthor{Danilkin, Kulikov}

\address{$^+$Institute of Theoretical and
 Experimental Physics, Moscow, Russia\\}
\address{$^*$Moscow Engineering
Physics Institute, Moscow, Russia\\}

\abstract{The possible description of Z(4430) as a pseudoresonance
structure in $\pi \psi'$ reaction, is considered. The analysis is
performed with single-scattering contribution to $\pi \psi'$
elastic scattering via $D^*D_1(2420)$ intermediate energy.}


\begin{document}
\maketitle

\section{Introduction}
The resonance structure with mass $M=4433\pm4\pm2$\,MeV and width
$\Gamma=45^{+18+30}_{-13-13}$ MeV in the charged quarkonium system
$\pi^{\pm}\psi'$ was found by the Belle Collaboration
\cite{Belle:2007}. On the other hand, BABAR Collaboration
\cite{BABAR:2008} did not see significant evidence for a
$Z(4430)^-$ signal in any of the processes investigated , neither
in the total $J/\psi \pi^-$  or $\psi(2S)\pi^-$   mass
distribution, nor in the corresponding distributions for the
regions  of $K\pi^-$ mass for which observation of the $Z(4430)^-$
signal was reported. Several mechanisms have been proposed to
explain the properties of the new resonance
\cite{Rosner:2007mu}-\cite{Bugg:2007vp}. In particular, Rosner
\cite{Rosner:2007mu} pointed out to the close by threshold of
$D_1(2420)\bar{D}^*(2010)$ state and suggested a mechanism of
production of $\pi\psi'$ in the decay $B\rightarrow KZ(4430)$,
$Z(4430)\rightarrow\pi^+\psi'$. The proximity of the threshold
invokes a possible near-threshold singularity, either due to a
pole of the amplitude (virtual or real loosely coupled bound state
of $D_1D^*$) \cite{Rosner:2007mu}-\cite{Meng:2007fu} or else due
to the threshold cusp \cite{Bugg:2007vp}.

In this letter we are trying to understand whether the Z(4430)
resonance can be due to pseudoresonance mechanism known for $\pi
d$ system \cite{Simonov_Velde:1978}. We analyze the structure of
the scattering amplitude for the reaction $\pi \psi'\rightarrow\pi
\psi'$  near the $D_1D^*$ threshold in the same way as was done
for $\pi d$ system near the $\triangle N$ resonance. It is well
known that the peak in the cross section for pion-nucleon ($\pi
N$) scattering around $T_{\pi}=180$ MeV is associated with the
$\triangle(1232)$. An analogous peak is observed in the cross
section for pion-deuteron ($\pi d$) scattering near $\triangle N$
threshold shifted slightly in position and broadened with respect
to the $\pi N$ peak (see Figure \ref{fig.1}). Therefore, one can
not exclude that the Z(4430) resonance, which lies near $D_1D^*$
threshold could be connected to the $D_1(2420)$ resonance, as it
takes place in the $\triangle(1232)$. The $D_1(2420)$ state with
mass $M=2420^{+1+2}_{-2-2}$ MeV and width
$\Gamma=20^{+6+3}_{-5-3}$ MeV was observed in
$D^{*\pm}(2010)\pi^{\mp}$ invariant distribution. Therefore the
dynamical picture of the pion charmonium scattering in our
approach is: the p-wave off-energy-shell charmonium decay to
$D^*\bar{D}^*$, then in the $\pi D^*$ scattering the creation of
$D_1(2420)$ resonance. The diagram corresponding to this reaction
is shown in Figure \ref{fig.2}.

In our paper, firstly, we calculate the scattering amplitude for
$\pi d$ system using a single Breit-Wigner resonance for
$\triangle(1232)$ and obtain a good description of $\triangle N$
resonance. Then we apply the same formulas to $\pi \psi'$
scattering in which the vertex of the $\psi'\rightarrow
D^*\bar{D}^*$ decay is calculated in the many channel formalism
developed in \cite{Simonov_Di-pion_decays:2007}. For simplicity
reasons we didn't include rescattering terms which slightly shift
the peak in the $\pi d$ case.

We pay a special attention to the influence of different
properties of deuteron and charmonium family. First of all its
different size: the deuteron is a large object with size
$R_d\sim4.3$ fm while charmonium $\psi'$ state has only
$R_{\psi'}\sim0.5$ fm. The analysis of the results and discussion
are given in the last section.

\section{The amplitude for $\pi d$ system}

For the sake of simplicity we neglect any spin dependence and
write the single-scattering non-relativistic term for the $\pi d$
amplitude:
\begin{equation}\label{M}
   M(\vec{k'},\vec{k})=\int\frac{d^3p}{(2\pi)^3}~\phi^*(\vec{p}-\frac{1}{2}\vec{k'})~M_{\pi
   N}(\vec{x'},\vec{x},W_1)~\phi(\vec{p}-\frac{1}{2}\vec{k})
\end{equation}
where
$\sqrt{s}=\sqrt{\vec{k}^2+m_{\pi}^2}+\sqrt{\vec{k}^2+m_{d}^2}$ is
the total invariant energy of the $\pi d$ system. In (\ref{M}) the
$\pi N$ amplitude depends on
\begin{eqnarray}\label{}\nonumber
&&\vec{x}=\vec{k}-\eta(p)\vec{p}\quad \vec{x'}=\vec{k'}-\eta(p)\vec{p}\\
\nonumber&&\eta(p)=\frac{\sqrt{\vec{p}^2+m_{\pi}^2}}{\sqrt{\vec{p}^2+m_{\pi}^2}+m_{N}}
\end{eqnarray}
and on the total invariant $\pi N$  energy $W_1$
\begin{equation}\label{}
W_1=\sqrt{\left(\sqrt{s}-\sqrt{\vec{p}^2+m_{N}^2}\right)^2-\vec{p}^2}.
\end{equation}
The $\pi N$ amplitude will be truncated to include only the
dominant resonance p wave in the following way:
\begin{equation}\label{}
M_{\pi N}=\frac{64}{3}\pi W_1 \vec{x'}~
\vec{x}\left(-\frac{\Gamma_R}{2q}\right)
\frac{1}{W_1-M_R+\frac{1}{2}i\Gamma_R}
\end{equation}
with momentum $q$ of the $\pi N$ system
\begin{equation}\label{}
q=\sqrt{\frac{(W_1^2-(m_{N}^2+m_{\pi}^2)^2)(W_1^2-(m_{N}^2-m_{\pi}^2)^2)}{4W_1^2}}.
\end{equation}
The deuteron wave function contains the deuteron pole:
\begin{equation}\label{}
\phi(\vec{p})=\frac{\sqrt{\alpha}}{(p^2+\alpha^2)(p^2+c^2)}
\end{equation}
with $\alpha=\sqrt{m_N\varepsilon_D}$, $\varepsilon_D$ being the
deuteron binding energy and $c\approx0.4$ GeV.
\begin{figure}
  \includegraphics[angle=270,width=0.45\textwidth]{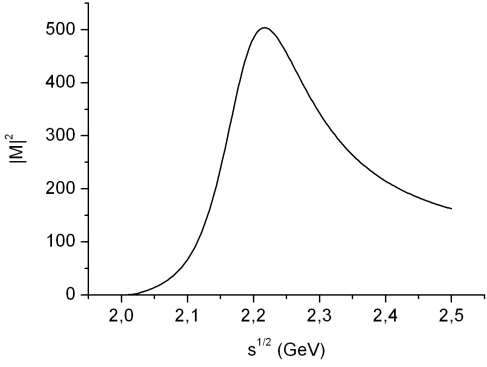}
  \caption{Figure 1. The squared $\pi$ $d$ scattering amplitude.}\label{fig.1}
\end{figure}

One can see in Figure \ref{fig.1} that the forward scattering
$(k=k')$ amplitude has a quite good resonance form which agrees
with the experiment result (see, for example
\cite{Thomas:1979xu}).

\begin{figure}
  \center{\includegraphics[angle=270,width=0.40\textwidth]{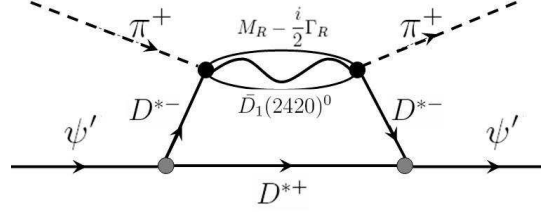}}
  \caption{Figure 2. Representation of the single scattering $\pi\psi'$ diagram.}\label{fig.2}
\end{figure}
\section{The amplitude for $\pi$ $\psi'$ system}

The $\phi(\vec{p})$ in (\ref{M}) for $\pi$ $\psi'$ system includes
the propogator and overlapped integral of the process
$\psi'\rightarrow D^*\bar{D}^*$
\begin{equation}\label{}
    \phi(\vec{p})=\frac{J(\vec{p})M_\omega}{E_{\psi'}-E_{D^*}-E_{\bar{D}^*}}=
    \frac{J(\vec{p})M_\omega}{M_{\psi'} - 2M_{D^*} - \frac{p^2}{M_{D^*}}}
\end{equation}
where  $J(\vec{p})$ is an overlapped matrix element between wave
functions $\Psi(nS)$ of the n-th charmonium state and $\psi(1S)$
of $D^*$($\bar{D}^*$) mesons states, which were derived in the
framework of many-channel formalism with decay channel coupling
\cite{Simonov_Di-pion_decays:2007}:
\begin{eqnarray}\label{J(p)}
    J(\vec{p})&=&\int \bar{y}_{123}\frac{d^3
    q}{(2\pi)^3}~\Psi(nS;c\vec{p}+\vec{q})~\psi(1S;\vec{q})~\psi(1S;\vec{q})
\end{eqnarray}
\begin{figure}
  \centering
  \includegraphics[angle=270,width=0.45\textwidth]{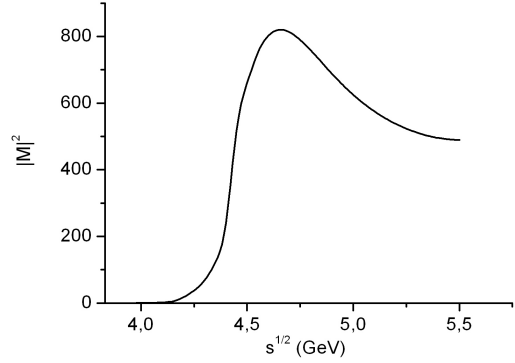}
  \caption{Figure 3. The squared $\pi$ $\psi'$ scattering amplitude.}\label{fig.3}
\end{figure}here $c=\frac{\Omega}{\Omega+\omega}$ ($\Omega$, $\omega$ is the
energy of heavy and light quarks in $D^*$ meson); $\bar{y}_{123}$
is defined by the Dirac traces of the amplitude given in appendix.
In eq. (\ref{J(p)}) $\Psi(nS), \psi(1S)$ are a series of
oscillator wave functions which are fitted to realistic wave
functions. We obtain them from the solution of the Relativistic
String Hamiltonian, described in \cite{Simonov_Dub_Kaid:1993fk},
\cite{Danilkin:2008bi}.

Figure \ref{fig.3} shows the squared $\pi$ $\psi'$ scattering
amplitude averaging over vector polarization
$\frac{1}{3}\sum\limits_{ii'}|M|^2$. As can be seen, the structure
has a too large width and a peak located near energy
$\sqrt{s}\sim4.7$ GeV and cannot be associated with Z(4430).

\section{Discussion}
\begin{figure}
  \centering
  \includegraphics[angle=270,width=0.45\textwidth]{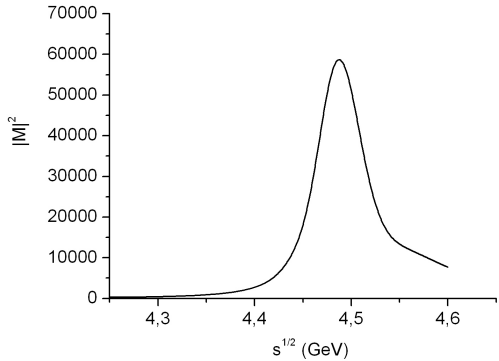}
  \caption{Figure 4. The squared $\pi$ $\psi'$ scattering amplitude in case of large charmonium size.}\label{fig.4}
\end{figure}

An important distinction between $\pi d$ and $\pi \psi'$ is the
difference in the deuteron and charmonium sizes, p wave decay
$\psi'$ to $D^*\bar{D}^*$ is also significant. It is interesting
that we can obtain a desirable resonance structure if $\psi'$ has
admixture of the near-threshold state with size $R\sim5$ fm due to
the coupling to the $D^*\bar{D}^*$ channel. In this case the width
turns out to be smaller $\Gamma\sim60$~MeV and the peak is shifted
to the position $\sqrt{s}\sim4.5$ GeV. This result is shown in
Figure \ref{fig.4}.

In our paper we have used dynamical picture of pion interaction
with heavy quarkonia corresponding to the diagram in Figure
\ref{fig.2}. Our analysis shows that there is no resonance near
$\sqrt{s}\sim4430$ energy in the $\psi'\pi$ system, unless an
admixture of large size near-the-threshold state is taken into
account.

We are grateful to Yu.A.Simonov for useful discussions. This work
is supported by the Grant NSh-4961.2008.2. One of the authors
(I.V.D.) is also supported by the grant of the {\it Dynasty
Foundation} and the {\it Russian Science Support Foundation}.


\section*{Appendix}
The vertex factor
$y_{123}=\frac{Z}{\sqrt{\bar{Z}_1\bar{Z}_2\bar{Z}_3}}$ is
calculated in the same way as in
\cite{Simonov_Di-pion_decays:2007}, namely from the Dirac trace of
the projection operators for the decay process, in our case this
is $\psi(nS)\rightarrow D^*\bar{D}^*$. Identify the creation
operators as $\bar{\psi}_c\gamma_i\psi_c$,
$\bar{\psi}_c\gamma_j\psi_d$, $\bar{\psi}_c\gamma_k\psi_d$ one has
for the decay process
\begin{eqnarray}\label{Z}
Z=tr(\gamma_i\Lambda^+\gamma_j\Lambda^-\Lambda^+\gamma_k\Lambda^-)
\end{eqnarray}
with the projection operators $\Lambda^{\pm}=\frac{m_k\pm
\omega_k\gamma_4\mp ip_i^{(k)}\gamma_i}{2\omega_k}$, k=c,d. Here
$\omega_k$ is the average energy of quark in given meson ($\Omega
=1.5$ GeV, $\omega =0.55$ GeV), $m_k$ is the pole mass of c and d
quarks ($m_c=1.4$, $m_d\approx0$). One can identifying the momenta
of $q,\bar{q},Q,\bar{Q}$ as in \cite{Simonov_Di-pion_decays:2007}:
\begin{eqnarray}\label{} &
\vec{p}_{\bar{q}}\text{=}-\vec{q}_1+\frac{\omega }{\omega
+\Omega}\,\vec{p}, &\quad \vec{p}_q\text{=}-\vec{q}_2-\frac{\omega
}{\omega +\Omega }\,\vec{p}
\\ \nonumber & \vec{p}_Q\text{=}\vec{p}-\vec{p}_{\bar{q}},&
\quad \vec{p}_{\bar{Q}}\text{=}-\vec{p}-\vec{p}_q
\end{eqnarray}
Finally one obtains from (\ref{Z}), taking into account that
$\vec{q}_2=-\vec{q}_1\equiv-\vec{q}$
\begin{eqnarray}\label{} \nonumber&&
Z=\frac{8 i m_Q }{16\omega^2\Omega^2}\{2 \omega  \Omega (\frac{p_k
\omega }{\omega +\Omega }-q_k) \mathbf{\delta_{i j}}+2 \omega
\Omega (\frac{p_j \omega }{\omega +\Omega }-q_j) \mathbf{\delta_{i
k}}
\\ &&+(p_i (-\frac{ \Omega  \omega ^2}{\omega +\Omega
}-\frac{p^2 \Omega \omega ^2}{(\omega +\Omega )^3}+\frac{2 \Omega
(p\cdot q) \omega }{(\omega +\Omega )^2}-\frac{q^2 \Omega }{\omega
+\Omega })
\\ \nonumber && +q_i (\omega ^2+2 \omega  \Omega
-q^2+\frac{2 \omega p\cdot q}{\omega +\Omega }-\frac{p^2 \omega
^2}{(\omega +\Omega )^2})) \mathbf{\delta_{j\, k}}\}
\end{eqnarray}

\end{document}